\documentclass[letterpaper]{aa}
\usepackage{txfonts}
\usepackage{graphicx}
\usepackage{subfigure}
\usepackage{natbib}

\bibpunct{(}{)}{;}{a}{}{,}

\begin{document}

\title{Lucky Imaging: High Angular Resolution Imaging in the Visible from the Ground\thanks{Based on observations made with the Nordic Optical Telescope, operated on the island of La Palma jointly by Denmark, Finland, Iceland, Norway, and Sweden, in the Spanish Observatorio del Roque de los Muchachos of the Instituto de Astrofisica de Canarias.}
}
\titlerunning{Lucky Imaging}
\author{N.M. Law \and C.D. Mackay \and J.E. Baldwin}
\institute{Institute of Astronomy, Cambridge, UK}

\date{Received - / Accepted -}

\abstract{
We use a Lucky Imaging system to obtain I-band images with improved angular resolution on a 2.5m telescope. We present results from a 10-night assessment campaign on the 2.56m Nordic Optical Telescope and quantify the performance of our system in seeings better than 1.0''. In good seeing we have acquired near diffraction-limited images; in poorer seeing the angular resolution has been routinely improved by factors of 2.5-4. The system can use guide stars as faint as I=16 with full performance and its useful field of view is consistently larger than 40" diameter. The technique shows promise for a number of science programmes, both galactic (eg. binary candidates, brown dwarfs, globular cluster cores) and extragalactic (eg. quasar host galaxies, damped Lyman-$\alpha$ absorbers).
}

\maketitle

\section{Introduction}

We have demonstrated in recent work the recovery of essentially diffraction limited I-band images with a novel imaging system on the 2.56m Nordic Optical Telescope (NOT) \citep{Baldwin_2001, Tubbs_2002, Tubbs_2003, Mackay_2004_Lucky}. Our Lucky Imaging system (LuckyCam) takes a sequence of images at $>$10 frames per second using a very low noise L3CCD based conventional camera. Each short exposure suffers different atmospheric turbulence effects, resulting in rapid variations in image quality. To construct a final image we select and co-add only those frames which meet a quality criterion. By varying the criterion we can trade off sensitivity against ultimate resolution. 

Lucky Imaging is a passive technique, so useful data is taken as soon as the telescope is pointed correctly. LuckyCam may be used in the same manner as any normal CCD camera system, with no special setup or operational requirements. 

Lucky Imaging shares the goal of Adaptive Optics (AO) systems - to enhance ground-based telescope resolution. AO systems are currently in use at several telescopes, and give very valuable resolution in the infrared. However, we are not aware of any systems routinely giving excellent resolution over a large ($>$5") field of view at wavelengths shorter than 1$\rm{\mu m}$.
 
In this paper we show that our prototype Lucky Imaging camera can give a very valuable increase in I-band resolution over a large area with only modest signal loss due to frame selection. Using data taken over 10 nights in 2000--2004 we explore the system's performance in I-band on a 2.5m telescope. This investigates a small part of the possible Lucky Imaging parameter space, which can be defined in terms of camera frame rate, passband and telescope size. 

In most observations presented here the camera is run approximately 3$\times$ too slowly to fully sample changes in the atmospheric turbulence and so we cannot demonstrate properly the full potential of Lucky Imaging. However, the very significant increase in resolution available with the current system is valuable for many science programmes.

  \begin{figure*}
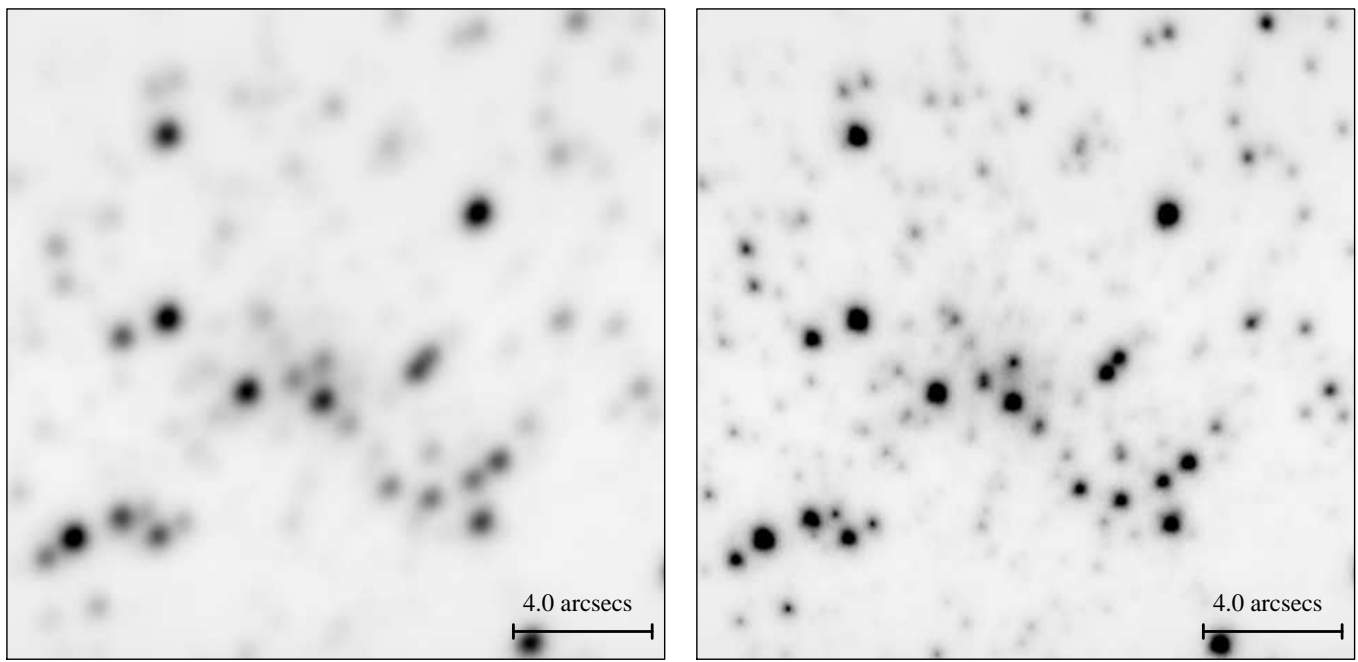

   \centering
   \resizebox{\textwidth}{!}
   {
	\subfigure{\includegraphics{autog_lowres.eps}}
	\subfigure{\includegraphics{10p2_lowres.eps}}
   }
   \caption{Lucky Imaging of the core of M15. The left panel is the output from an autoguider system (FWHM 0.63''); the right shows the 10\% frame selection Lucky output from the same exposures (FWHM 0.26''). Both images are on the same grayscale. The increased resolution makes more faint objects visible in the Lucky image.}
   \label{FIG:Example_field}
   \end{figure*}

In section \ref{SEC:Method} we describe our camera and data reduction techniques. In section \ref{SEC:PSF} we describe the point spread function improvements produced and the specific data reduction techniques we use. In section \ref{SEC:Lucky_perform} we detail the results of Lucky Imaging trials to date. Section \ref{SEC:SN} addresses the effect on limiting magnitudes of the Lucky Imaging frame selection; section \ref{SEC:Lucky_reqs} discusses the guide star and hardware requirements.

\section{The Lucky Imaging Technique}

\label{SEC:Method}

The Cambridge Lucky Imaging system (LuckyCam) is based on an E2V Technologies L3CCD read out with a 4 MHz pixel rate and mounted at the focus of a simple reimaging camera. The on-chip gain stage of the L3CCD raises the signal from incoming light sufficiently to allow individual photons to be detected with good signal to noise, even at high frame rates - for more details see e.g. \citet{Mackay_2003, Basden_PC, Mackay_2004_L3CCD}. 

Typically the camera is run in full-frame (552x512 pixel) 12 frames per second (FPS) mode, although the frame size may be reduced for higher frame rates. The image scale is 0.04"/pixel, giving a field of view of 22$\times$20.5 $\rm{{arcseconds}}$. This slightly undersamples the 0.08" FWHM I-band diffraction limited PSF that would be produced by the Nordic Optical Telescope.

The frame selection algorithm, implemented (currently) as a post-processing step, is summarised below:

\begin{enumerate}
\item{A Point Spread Function (PSF) guide star is selected as a reference to the turbulence induced blurring of each frame.}
\item{The guide star image in each frame is sinc-resampled by a factor of 4 to give a sub-pixel estimate of the position of the brightest speckle.}
\item{A quality factor (currently the fraction of light concentrated in the brightest pixel of the PSF) is calculated for each frame.}
\item{A fraction of the frames are then selected according to their quality factors. The fraction is chosen to optimise the tradeoff between the resolution and the target signal-to-noise ratio required.}
\item{The selected frames are shifted-and-added to align their brightest speckle positions.}
\end{enumerate}

 \begin{figure*}
   \centering
   \resizebox{\textwidth}{!}
   {
	\subfigure{\label{FIG:Prof_good}\includegraphics{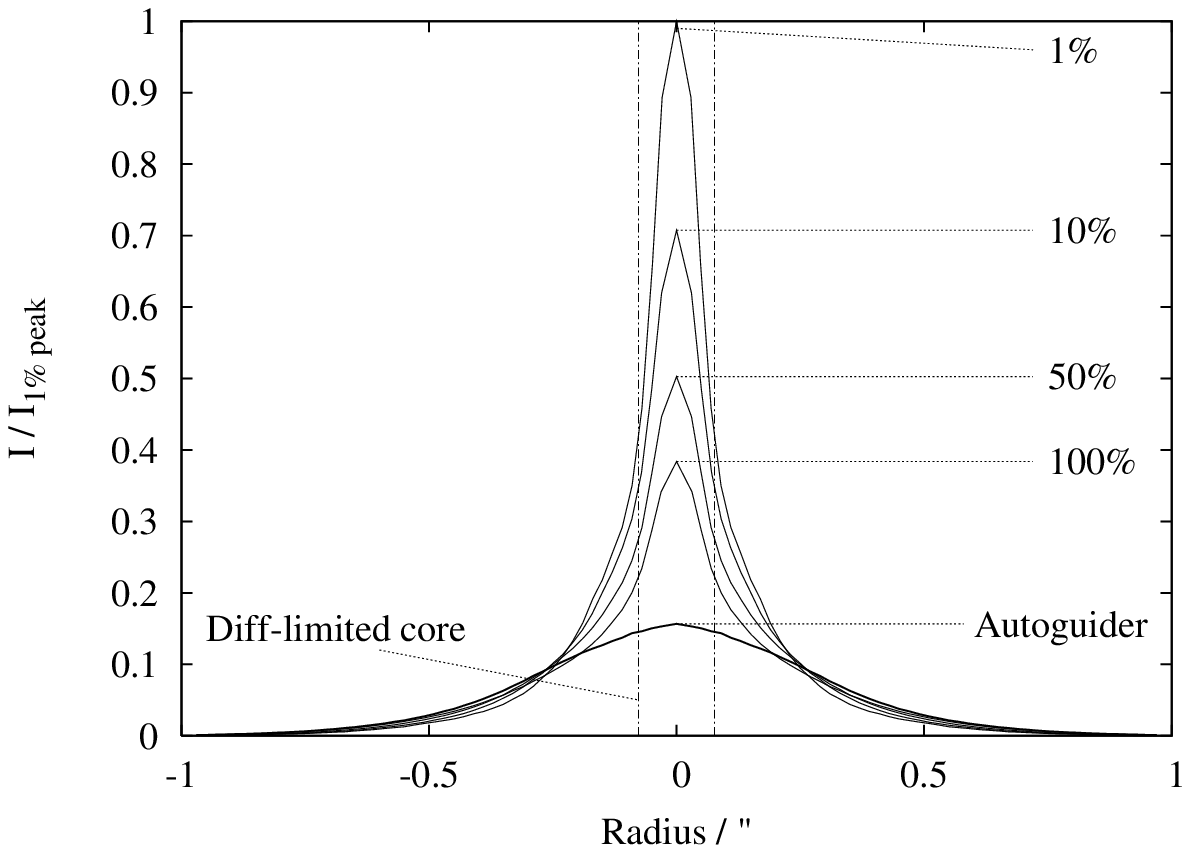}}
	\subfigure{\label{FIG:Prof_bad}\includegraphics{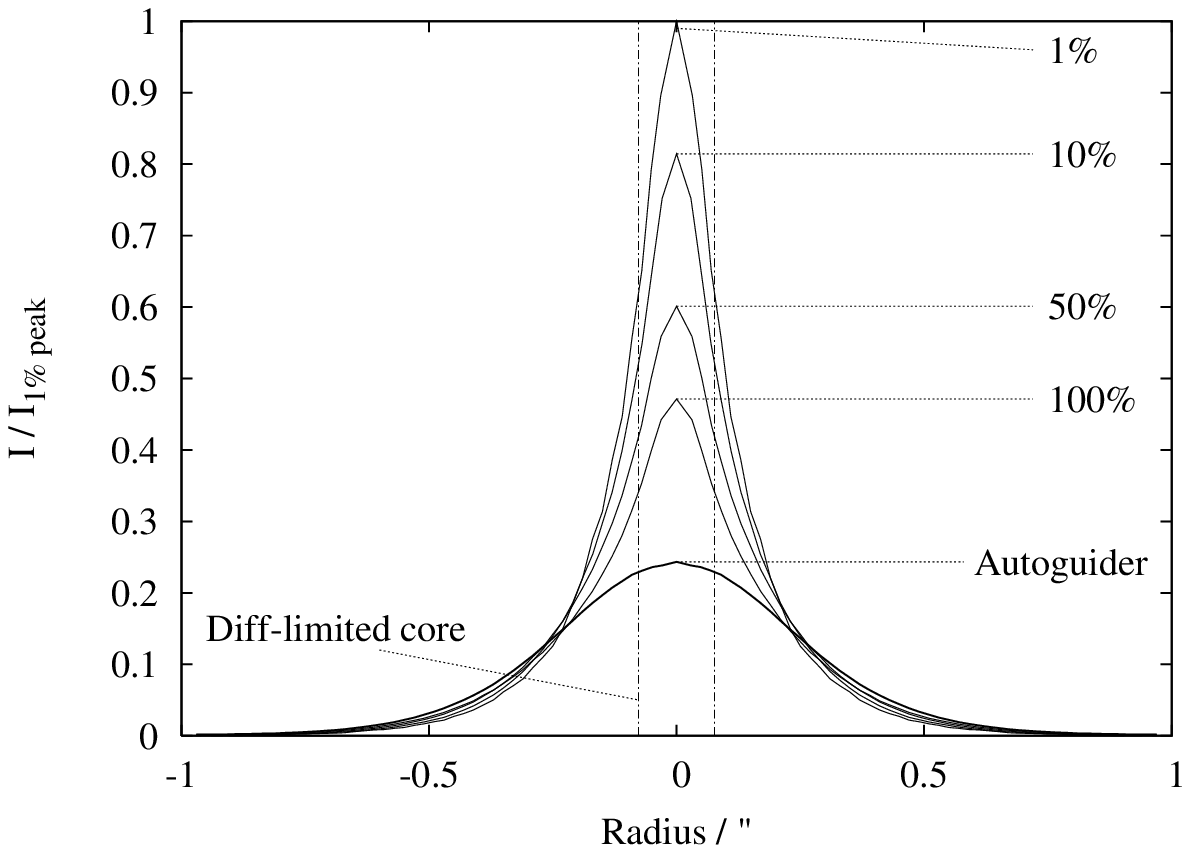}}
   }
      \caption{Average radial Lucky Imaging profiles. Data for the left figure was taken at 30Hz in 0.6'' seeing; the right figure shows the results of 12Hz imaging in 0.55'' seeing. 100\% selection is simple shift-and-add. When selecting 1\% of frames the Strehl ratio is more than doubled relative to the 100\% selection, while the light from the star is concentrated into an area approximately four times smaller.}
         \label{FIG:Profiles}
   \end{figure*}

Faint guide star PSFs are affected by photon shot noise, which can lead to an image being positioned on the basis of a noise spike rather than an actual bright speckle. This noise is reduced by convolving the faint reference star image with a theoretical diffraction-limited PSF.

An implementation of the Drizzle algorithm \citep{Drizzle} is used for the image alignment step. The algorithm resamples the images by a factor of two and minimises the information lost by our slightly undersampled pixel scale. Drizzle is especially suitable as we are summing many frames displaced by random non-integer pixel offsets.

Figure \ref{FIG:Example_field} gives a typical example of the improvements obtained with Lucky Imaging.

\section{The PSFs produced by Lucky Imaging}

\label{SEC:PSF}
Examples of the general form of the Lucky Imaging PSF are given in fig. \ref{FIG:Profiles}. As with adaptive optics images, the radial shape takes the expected form for an image with partially compensated Kolmogorov turbulence (e.g.. \citet{Hardy_book} and references therein) - a wide halo and a central compact core. As the selection of frames is made more stringent the fraction of light in the compact central core progressively increases.
   
   \begin{figure*}
   \centering
   \resizebox{\textwidth}{!}
   {\includegraphics[width=\textwidth]{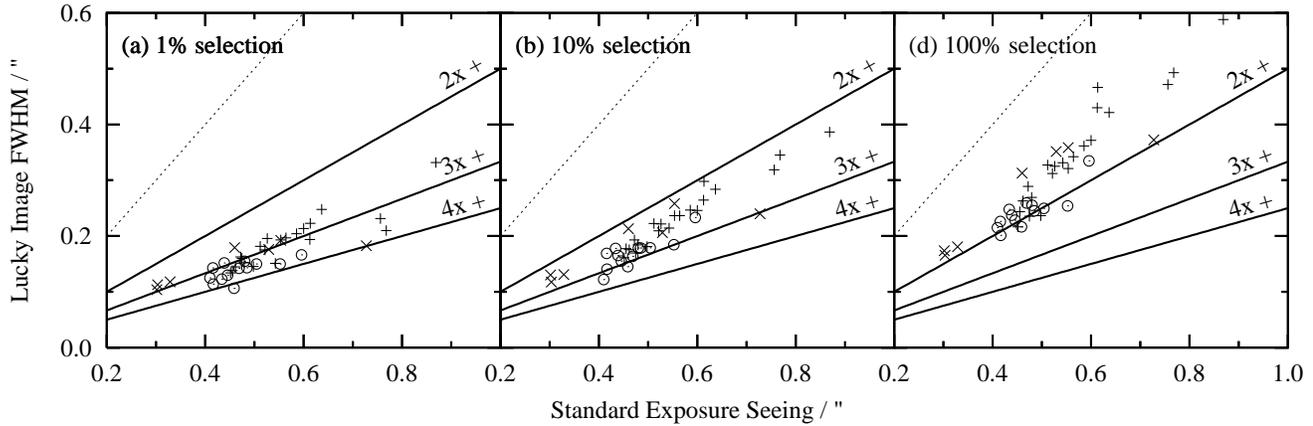}}
      \caption{FWHM of target stars 1-3" from the reference star. Lines correspond to resolution increase factors of $1\times$, $2\times$, $3\times$ \& $4\times$. Most of the scatter in the FWHMs is caused by the range of target star to guide star distances. Vertical crosses are runs taken at 12Hz in 2003 \& 2004; diagonal crosses are runs taken at 18Hz in 2001 \& 2002 and circles are runs taken at 36Hz in 2003.}
         \label{FIG:FWHM_3_plots}
   \end{figure*}

The Strehl ratio (commonly used as a high-resolution imaging performance measurement) is the peak value of a PSF divided by the theoretical diffraction-limited value. Figure \ref{FIG:Profiles} clearly shows that Lucky imaging offers a substantial Strehl ratio improvement over both autoguider and shift-and-add (effectively 100\% selection)  systems. We have routinely measured I-band Strehl ratios in the range of 0.15-0.2 at high frame rates in good seeing on the 2.5m NOT.

We note that the NOT is not designed or calibrated to produce diffraction limited images. Although the random phase variations in the atmosphere can compensate for telescope mirror (or focusing) errors they do so at a lower probability than the production of a good image with a perfect telescope. In the case of the NOT small scale mirror irregularities limit the peak Strehl in the absence of an atmosphere to around 0.2. Optics designed for high Strehl ratios would increase the probability of obtaining a high quality frame substantially \citep{Tubbs_thesis}. 

We cannot directly compare the Lucky Imaging performance to a standard imaging system as we are not able to take images simultaneously with the two systems. Rapidly changing seeing can then lead to a bias. To avoid the bias in this paper we apply a simulation of the NOT autoguider system to our LuckyCam data, effectively giving a simultaneous test of the two systems.

The quoted seeings are measured from the FWHM of 2D Moffat profile fits to simulated autoguider images and agree with those measured at intervals by conventional means during the observations. They are thus measured at an effective wavelength of $\rm{\sim850 nm}$ and should be increased by approximately 10\% to convert to a standard seeing quoted at 550nm. All other FWHMs detailed in this paper are similarly measured from 2D Moffat profile fits.

\section{Lucky Imaging Performance}
\label{SEC:Lucky_perform}

   \begin{figure*}
   \centering
   \resizebox{\textwidth}{!}
   {
	\subfigure{\includegraphics{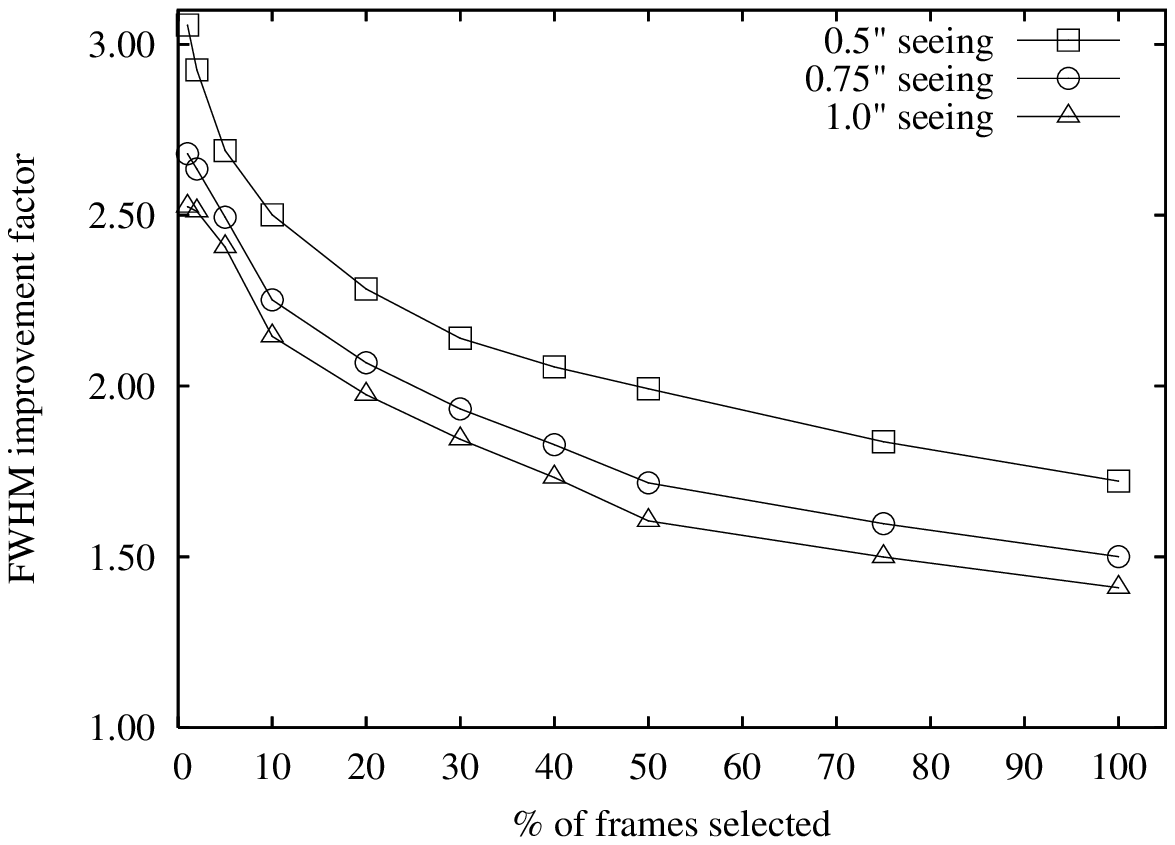}}
	\subfigure{\includegraphics{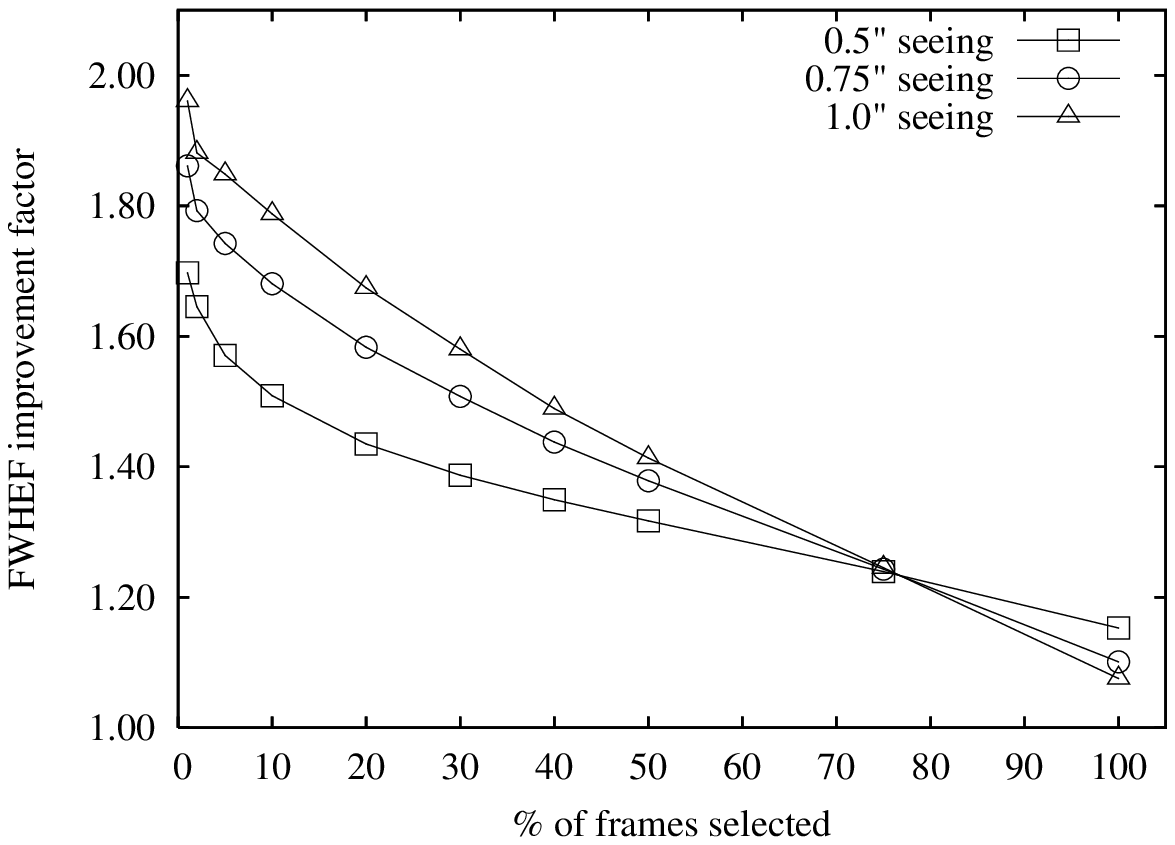}}
   }
      \caption{The factors by which FWHM and FWHEF (full width at half enclosed flux) are improved at slow frame rates (12FPS), at three seeings. Obtained from linear fits to figure \ref{FIG:FWHM_3_plots} and similarly derived results for the FWHEF. Starting at the leftmost point of each line, datapoints correspond to selecting frames at the 1\%, 2\%, 5\%, 10\%, 20\%, 30\%, 40\%, 50\%, 75\% and 100\% levels respectively.}
         \label{FIG:Improved}
   \end{figure*} 

The present results are based on observations taken during 10 nights on five observing runs in May, June \& July, 2000--2004. The 2.56m Nordic Optical Telescope on La Palma in the Canary Islands was used for all observations, which were principally made in several different fields in the cores of the Globular Clusters M3, M13 and M15. The 2001-2004 dataset is split into 42 short (2-3 minute) runs, totalling $\rm{\sim}$132,000 frames. All fields were observed in I-band.
 
We here investigate science-target performance - i.e. that for a star near the guide star. In each of the fields we chose a target star for PSF measurement at 1-3'' separation from the guide star.

The improvement in FWHM obtainable in different atmospheric conditions and with different percentage selections is shown in fig. \ref{FIG:FWHM_3_plots}. Much of the scatter in values is due to the range of distances of target stars from the reference star (section \ref{SEC:Isoplan}).

\label{Slow_frame_rate}

Figure \ref{FIG:FWHM_3_plots} shows that under a wide range of conditions the resolution is improved by factors as large as $\times$4 in the most stringent selections in all of the 42 observations reduced. Less stringent selections give smaller improvement. There is an approximately linear correspondence between the autoguided long exposure seeing and the resolution attainable. This suggests that, at least over the $\sim$200 second timescale of these observations, the standard measures of seeing are a reasonable guide to the atmospheric turbulence statistics. We can thus adopt the seeing as our standard measure of the atmospheric turbulence strength in a particular run.

Figure \ref{FIG:Improved} shows the effects of selecting differing fractions of images using empirical fits to the 12FPS data. Although limited by the slow frame rate, in 0.5" seeing the most stringently selected images are within $\sim$0.06'' of the 0.08" diffraction limit and Strehl ratios are $>$0.1.

The advantages of the full-frame field of view led us to obtain most of our data in LuckyCam's relatively slow 12 or 18 FPS modes. To obtain images with light reliably concentrated into bright single speckles, unblurred by image motion, we must oversample the atmospheric coherence time. This would typically require $>$ 40 FPS at the NOT. However, the image quality varies on every timescale slower than the coherence time and so image selection can always be expected to improve the resolution - even if we use very slow frame rates.

To acquire a high-quality frame we require both an excellent point spread function and a stable atmosphere for longer than our frame integration time; this occurs with increasing probability as we increase the frame rate. As expected, the faster frame rates presented in fig. \ref{FIG:FWHM_3_plots} on average give a greater resolution improvement than the slower rates. In good seeing the output resolution at fast frame rate is within 0.03" of the diffraction limit of the telescope, limited at least in part by LuckyCam's slight PSF undersampling.

     \begin{figure}
   \centering
   \resizebox{\columnwidth}{!}
   {\includegraphics{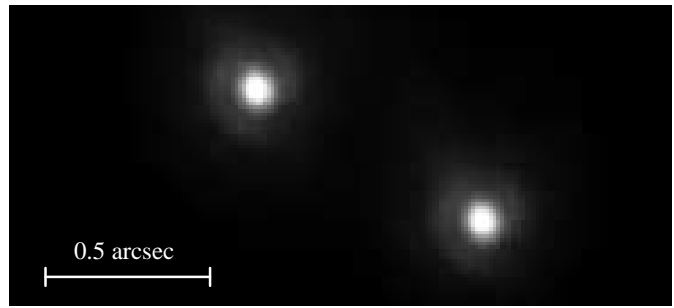}}
      \caption{A 1\% selected image of $\rm{\zeta}$ Bo\"otis with a Strehl ratio of 0.26 and a FWHM of $\sim$0.1", taken in 0.42" seeing. }
         \label{FIG:Airy_rings}
   \end{figure}

Although the resolution can be adjusted by altering the number of frames selected, in poorer seeing the probability of a superb frame is greatly reduced. Empirically this appears to limit the LuckyCam resolution increase to a factor of $\sim$3-4 when the seeing is poor.

The FWHM gives a good estimate of the system's performance for resolution enhancement - i.e. the separation of two closely separated objects. However, for many observations (notably precision photometry) the degree of light concentration within the larger halo is also important. Figure \ref{FIG:Improved} also details the Full Width at Half Enclosed Flux (FWHEF) performance. The FWHEF is the aperture size which contains half the light of the star and thus is a better measure for crowded observations where photometry is to be performed. The current LuckyCam improves the FWHEF by up to a factor of two, corresponding to concentrating half the light from a star into an area four times smaller than in seeing-limited images.

 With a faster camera higher resolutions have been reached. Figure \ref{FIG:Airy_rings} shows an I-band image of the 0.8" binary $\rm{\zeta}$ Bo\"otis with a Strehl ratio of 0.26, taken in 0.42" seeing. The first Airy ring is clearly visible at a radius of $\sim$0.1", showing the point spread function is indeed diffraction-limited in width. This dataset was taken at 200Hz, approximately $5\times$ faster than the coherence time of the atmosphere requires.

\subsection{Isoplanatic Patch}
\label{SEC:Isoplan}
Resolution enhancement systems such as LuckyCam provide useful turbulence correction only within some angular distance to the guide star. This distance is commonly quantified by the isoplanatic patch radius - the radius at which the Strehl ratio of stars is reduced by a factor of $e^{-1}$ from those very close to the guide star. We here use this definition of our useful field size, although noting our low frame rate wide-field data has generally smaller Strehl ratios (of order 0.1) than are usually used for this measurement. 

\begin{figure}
  \centering
   \resizebox{\columnwidth}{!}
   {\includegraphics[width=\columnwidth]{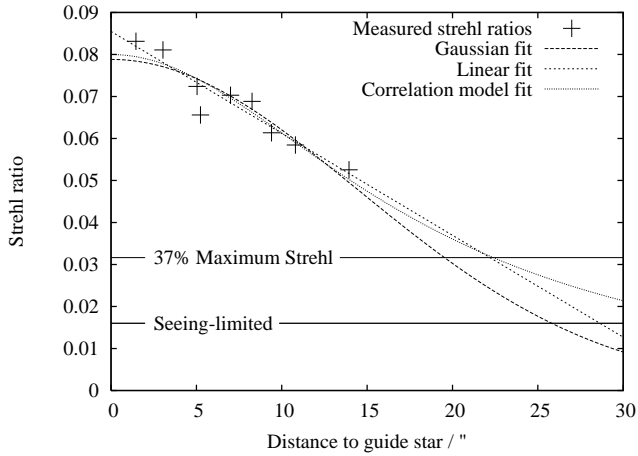}}
      \caption{An example set of isoplanatic patch extrapolations. Measured Strehl ratios of a number of stars (from a 0.5" seeing run at 1\% selection) are plotted as a function of distance from the reference star; three empirical models are fit.}
         \label{FIG:Isofit}
\end{figure}

The difference in Strehl ratio between stars close to the guide star and those up to 25" off-axis (our field limit) can be measured for several of the globular cluster fields in our dataset. The isoplanatic patch radius is then calculated from the point at which an empirical function fitted to the Strehl ratios drops to $e^{-1}$ of its on-axis value. The models (fig. \ref{FIG:Isofit}) suggest that in five fields of M15 taken over two nights in 2003 the isoplanatic patch size ranged from 17-30" in radius. Selecting a larger fraction of the frames gives an improved patch size (up to 10" wider) as the degree of turbulence correction achieved on-axis is decreased along with the output resolution.

It is clear that the atmospheric turbulence corrections produced even with a low frame rate camera have a remarkably large effective radius, in most cases larger than LuckyCam's field of view. These results (from data taken in 2003) agree with earlier Lucky Imaging results presented in \cite{Tubbs_2002}, suggesting that the large patch size may be inherent to the technique over a wide range of conditions.

\begin{figure}
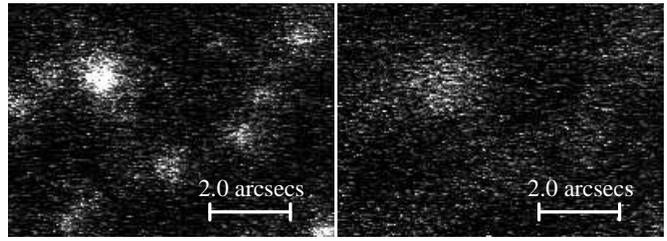

   \centering
   \resizebox{\columnwidth}{!}
   {
	\subfigure{\includegraphics{good_psf.eps}}
	\subfigure{\includegraphics{bad_psf.eps}}
   }
      \caption{An example of fast seeing variations. These images were taken with an 83ms exposure time in the core of M3 in average 0.85" seeing - \emph{0.33 seconds} apart. This is not a results of the short timescale statistical fluctuations in image resolution expected for a constant $\rm{r_0}$ - the 5-second averaged seeing changed by approximately a factor of two between these two images. Although extreme, these fast variations are present in most of the runs analysed here. Each bright pixel in these images is a detection of at least one photon.}
         \label{FIG:Good_Bad_PSFs}
   \end{figure}

The isoplanatic patch for speckle imaging is expected to be proportional to $r_0$, the standard atmospheric coherence length (\citet{Vernin_1994} and references therein). $r_0$ is known to vary on relatively short timescales (fig. \ref{FIG:Good_Bad_PSFs}). If LuckyCam selects periods when $r_0$ is larger (and not just periods when the turbulence induced phase errors randomly sum to a small value) it would be expected that those periods would have larger isoplanatic patches, as we observe.

  \begin{figure}
   \centering
  \resizebox{\columnwidth}{!}
    {\includegraphics[width=\columnwidth]{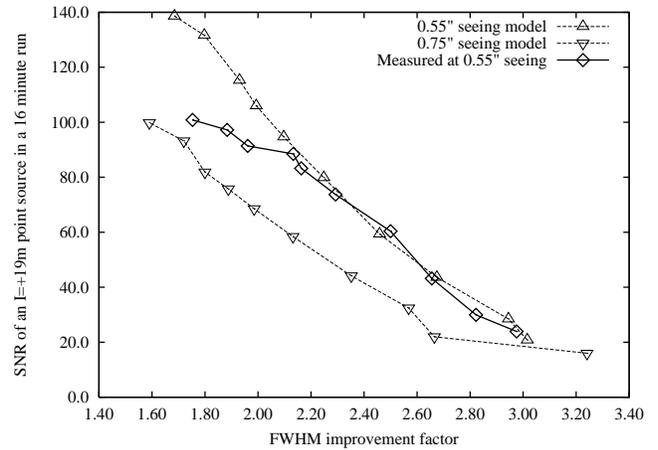}}
     \caption{The SNR (both measured and modelled) of an I=+19m point source in a 16 minute exposure as a function of the Lucky Imaging improvement in FWHM. The aperture size is set to the PSF diameter at half enclosed flux. Starting at 100\% at the upper end of each line the data points represent frame selection fractions of 100\%, 75\%, 50\%, 40\%, 30\%, 20\%, 10\%, 5\%, 2\% and 1\%. Note that these observations were made in the central core of M15 and have $\sim4\times$ higher background than an empty field - i.e. the SNRs are reduced by a factor of two.}
         \label{FIG:SN_Sel}
 \end{figure}

\section{Signal-to-noise Ratio (SNR) Considerations}
\label{SEC:SN}
An improvement in resolution allows smaller photometric apertures to be used. A smaller aperture contains fewer pixels and so less sky noise and (here negligible) detector noise, giving an increased SNR.

In fig. \ref{FIG:SN_Sel} we compare modelled LuckyCam SNR performance with optimised apertures (based the performance detailed in section \ref{Slow_frame_rate}) to measured results for an I=+19m star in the core of M15. As for this star the noise in these images is background dominated, the SNR is measured by comparing the signal contained inside an aperture to the RMS noise in a nearby empty (background) aperture.

The 0.55" seeing model agrees well with the measured SNR in high-resolution observations but diverges by $\sim$25\% at lower resolutions. This is due to increased crowding (our fields are in the cores of globular clusters) leading to an increased background. This aside, the standard CCD photometric noise estimates model LuckyCam's performance well, once the applicable aperture size changes are taken into account.

Although a 1\% selection of frames would be expected to decrease the SNR by a factor of $\sqrt{100} = 10$ for a star in an empty field, the smaller aperture sizes that can be used reduce the decrease to $\sim6.5\times$ in both seeings in fig. \ref{FIG:SN_Sel} . At the 50-75\% selection level the SNR matches that available in a standard autoguided image, with resolution increased by a factor of $\sim2$.

We emphasise that Lucky Imaging's ability to increase resolution by large factors even in poor seeing effectively increases the usable observation time at the telescope. For example, if a particular observation requires 0.5" seeing selecting a few percent of frames in 1.5" seeing will produce usable data in telescope time that would otherwise have been lost.

\section{Lucky Imaging Requirements}
\label{SEC:Lucky_reqs}
In this section we detail the two main requirements for a successful Lucky Imaging observation - a sufficiently bright guide star and a sufficiently fast camera.

   \subsection{Reference Star Flux}
   \label{SEC:Ref_star_flux}

   \begin{figure}
   \centering
   \resizebox{3in}{!}
   {\includegraphics[width=3in]{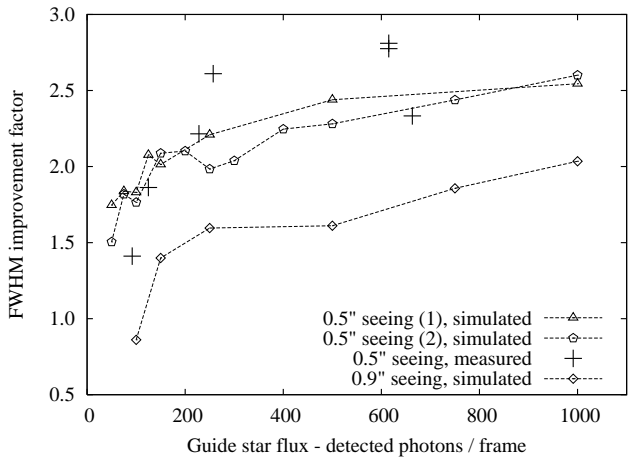}}
      \caption{The factor by which the target star FWHM is improved as a function of the guide star flux at a 10\% selection. The dashed lines show results based on simulating a faint star by adding photon \& L3CCD multiplication noise to a bright star's PSF; also shown (with crosses) are measurements using actual stars.}
         \label{FIG:Ref_flux_seeings}
   \end{figure}

We simulate the effect of using a faint guide star by rescaling a bright ($>$ 2000 photons / frame) star PSF to a specified flux, taking Poisson and L3CCD multiplication noise into account. As a check of the simulations' accuracy we also performed the same measurements with some non-simulated faint guide stars.

Figure \ref{FIG:Ref_flux_seeings} shows, at a 10\% selection, the reference star flux requirements at two seeings. The requirements are similar at all percentage selections.

At a guide star flux of 150 photons/frame ($\rm{I\sim+16m}$) the resolution is reduced by 22\% from that achieved with a very bright star. Because there appears to be a rapid falloff in resolution below that flux level we adopt 150 photons/frame as the minimum flux required for high resolution imaging.

The more spread out PSFs in poorer seeing require more photons to give an acceptable SNR. The real (non flux rescaled) guide stars give matching results, although with higher noise induced by anisoplanatism.

The $\sim$150 photons/frame requirement can be understood in the following way. Even in poor seeing the best frames consist of a single speckle surrounded by a halo. If the speckle contains only a few tens of photons, Poisson shot-noise becomes a limiting factor for frame selection on the basis of Strehl ratios - thus degrading the output resolution.

If a field contains several stars which are individually fainter than the guide star limit we have found it is possible to add the stars' PSFs together to provide a useful estimate of a bright guide star PSF, which can then be used in the standard way for Lucky Imaging.

\subsection{Sky Coverage}

The isoplanatic patch and reference star flux requirements directly give us LuckyCam's achievable sky coverage. If the system is to achieve its most useful resolution gains we require an $\rm{I\sim16}$ reference star within 25'' of our target.

At a galactic latitude of $\rm{60-70^o}$ approximately 7\% of the sky can be used with chance-placed guide stars, based on averaged star counts from the USNO-B1.0 catalogue \citep{USNO}. 30\% of the sky is accessible at a latitude of $\rm{20-30^o}$  while at lower galactic latitudes sky coverage is virtually complete.

\subsection{Frame rates}
\label{SEC:Frame_rates}

   \begin{figure}
   \centering
   \resizebox{\columnwidth}{!}
   {\includegraphics[width=\columnwidth]{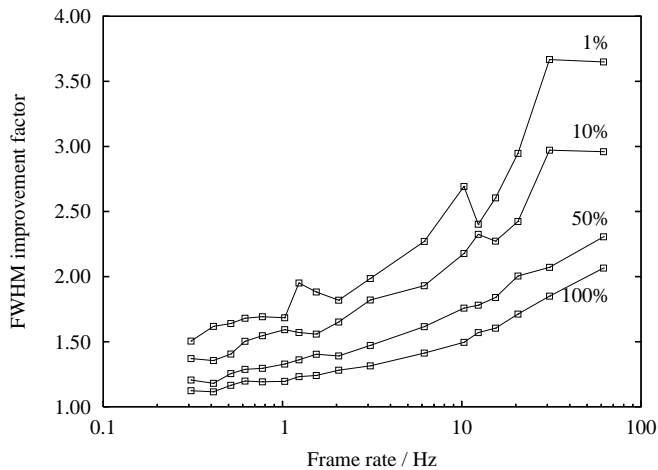}}
      \caption{Imaging performance as a function of frame rate for in 0.42'' seeing. Shift-and-add (100\% selection) FWHM performance is always a significant improvement over the autoguider performance because the autoguider uses the image centroid rather than the PSF peak as a position reference.}
         \label{FIG:Exp_time}
   \end{figure}

Freezing the effects of atmospheric turbulence requires exposures that are shorter than the atmospheric coherence time. One sufficiently fast (60Hz) trial dataset was taken 2003 and can be temporally rebinned to produce a range of effective effective frame rates (fig. \ref{FIG:Exp_time}). Although this is a single observation the performance at 12FPS and 30FPS is consistent with that found in our earlier more numerous trials and is thus probably not atypical.

Selection of frames (as opposed to simple shift-and-add) gives an improvement at all measured frame rates. At even 0.3 FPS the frame selection gives gains in FWHM of over 1.6$\times$, with an estimated limiting reference star magnitude of I=+19m in dark time.

At $\sim$30 FPS the atmospheric coherence time is well sampled in the best 10\% of frames; there is no improvement with faster frame rates. The frames with poorer quality do, however, benefit from still faster rates. The periods of poor seeing have a relatively smaller atmospheric coherence time because there are more isophase patches per unit area to be swept over the telescope aperture by the wind. The more inclusive frame selections include these periods and so require a faster frame rate than when selecting only the very best frames. 

Running at only 12FPS the current system cannot properly demonstrate the full potential of Lucky imaging; we have tested a new version of LuckyCam for science use in Summer 2005 that provides the faster full-field frame rates required for full performance.

\section{Conclusions}
\label{SEC: concs}

Lucky Imaging offers reliable and very valuable resolution enhancement in all encountered conditions, even with our slow non-ideal camera. In poor seeing FWHMs are improved by up to a factor of four; in good seeing near diffraction limited images are achieved. The angular radius over which the system can usefully operate in I-band is as large or larger than our 22''x20.5'' field of view in all measured fields. The current system can work with guide stars as faint as $\rm{I=16}$ with full performance, giving near 100\% sky coverage at galactic latitudes $<\rm{20^o}$ and $>$7\% sky coverage at $<\rm{70^o}$. Signal-to-noise ratio losses from frame selection are compensated by the greatly improved resolution. 

The system increases the range of conditions that are usable for  particular observations - and can thus use otherwise lost telescope time, as well as giving resolutions in the visible far in excess of those of conventional CCD cameras on ground-based telescopes.

Science programmes targetting both galactic (eg. binary candidates, brown dwarfs, globular cluster cores) and extragalactic (eg. quasar host galaxies, damped Lyman-$\alpha$ absorbers) objects will benefit greatly from uncomplicated, reliable and cheap high resolution I-band imaging.

\appendix

\begin{acknowledgements}
The authors would like to particularly thank Graham Cox at the NOT for several helpful discussions, and are very grateful to many other helpful NOT staff members. NML acknowledges support from the UK Particle Physics and Astronomy Research Council (PPARC).
\end{acknowledgements}

\bibliographystyle{aa}
\bibliography{nml_paper}

\end{document}